\documentclass[11pt]{article}
\usepackage{graphicx}
\usepackage{amssymb}
\usepackage{latexsym}

\newcommand{\be}{\begin{equation}}
\newcommand{\ee}{\end{equation}}
\newcommand{\bea}{\begin{eqnarray}}
\newcommand{\eea}{\end{eqnarray}}
\newcommand{\comment}[1]{}
\newcommand{\lsim}{\!\mathrel{\hbox{\rlap{\lower.55ex \hbox{$\sim$}} \kern-.34em \raise.4ex \hbox{$<$}}}}
\newcommand{\gsim}{\!\mathrel{\hbox{\rlap{\lower.55ex \hbox{$\sim$}} \kern-.34em \raise.4ex \hbox{$>$}}}}

\newcommand{\GeV}{~\mathrm{GeV}}
\newcommand{\MeV}{~\mathrm{MeV}}
\newcommand{\keV}{~\mathrm{keV}}
\newcommand{\eV}{~\mathrm{eV}}
\newcommand{\tabwide}{\rule[-3ex]{0pt}{8ex}}

\begin{document}

\begin{titlepage}

\begin{flushright}
HUTP-04/A033
\end{flushright}

\vspace{.5in}

\begin{center}

{\LARGE \bf Universal Dynamics of Spontaneous Lorentz Violation \\ \medskip and a New Spin-Dependent Inverse-Square Law Force}

\vspace{.3in}

{\Large Nima Arkani-Hamed, Hsin-Chia Cheng, Markus Luty and Jesse Thaler}

\vspace{.15in}

{\large \it Jefferson Physical Laboratory, Harvard University, Cambridge, MA 02138}

\vspace{.15in}

\end{center}

\begin{abstract}

We study the universal low-energy dynamics associated with the
spontaneous breaking of Lorentz invariance down to spatial rotations.
The effective Lagrangian for the associated Goldstone field can be
uniquely determined by the non-linear realization of a broken time
diffeomorphism symmetry, up to some overall mass scales. It has
previously been shown that this symmetry breaking pattern gives rise to a Higgs phase of gravity,
in which gravity is modified in the infrared. In this paper, we study the effects of direct couplings between the Goldstone boson and standard model fermions, which necessarily accompany
Lorentz-violating terms in the theory. The leading interaction is
the coupling to the axial vector current, which reduces to spin in
the non-relativistic limit. A spin moving relative to the ``ether"
rest frame will emit Goldstone \v{C}erenkov radiation. The
Goldstone also induces a long-range inverse-square law force between
spin sources with a striking angular dependence, reflecting the
underlying Goldstone shockwaves and providing a smoking gun for this
theory. We discuss the regime of validity of the effective theory
describing these phenomena, and the possibility of probing
Lorentz violations through Goldstone boson signals in a way that is
complementary to direct tests in some regions of parameter space.

\end{abstract}
\thispagestyle{empty}
\end{titlepage}

\pagebreak

\section{Introduction}

Ever since Einstein banished the ``luminiferous ether" from
physics in 1905, Lorentz invariance has been a fundamental part of
our description of nature. However, there has been growing
interest in studying possible violations of this symmetry. One
reason is that Lorentz invariance is clearly broken on large
scales by cosmology, which does provide us with a preferred frame, namely
the frame in which the Cosmic Microwave Background Radiation (CMBR) is spatially isotropic, and to a good first approximation, Hubble friction
provides a dynamical explanation for why all observers come to
rest in this frame.  But we normally
don't think of the CMBR or other cosmological fluids as
``spontaneously breaking" Lorentz invariance --- we instead think of
them as excited states of a theory with a Lorentz invariant ground
state. Indeed, the expanding universe redshifts away the CMBR,
till in the deep future the universe is empty and Lorentz
invariance (or de Sitter invariance) is recovered. Also, Lorentz invariance is certainly a good symmetry locally on scales much
smaller than the cosmological horizon.

It is therefore interesting to instead explore the possibility
that Lorentz invariance is broken in the vacuum, so that even as
the universe expands and we asymptote to flat space or de Sitter
space, local Lorentz violations still remains. There is a large
literature on studying Lorentz-violating extensions of the
standard model (see \emph{e.g.}\ Refs.~\cite{Colladay:1998fq, Colladay:1996iz, Coleman:1998ti, LorentzOne, LorentzTwo,Bluhm:2003ne}).
These works mainly concentrate on putting experimental limits on
the coefficients of Lorentz-violating operators in the theory.
However, we cannot have spontaneously broken Lorentz invariance without having
additional {\it dynamical} effects. The reason is that there must
be Goldstone bosons associated with the breaking of Lorentz
invariance. Thus, associated with every Lorentz-violating operator we
have a coupling to the Goldstone boson, leading to interesting new
physical effects that we will examine in this paper.

We will focus on the simplest pattern for the breaking of Lorentz
invariance which leaves rotations intact. There is then a
preferred ``ether" rest frame, where this breaking can be thought
of as a non-zero vacuum expectation value (vev) for the time component of a vector field
$A_\mu$, $\langle A_0 \rangle \neq 0$. In fact, the minimal number
of new degrees of freedom required for this symmetry breaking
pattern arises if we take $A_\mu = \partial_\mu \phi$, as was done
in Ref.~\cite{GhostOriginal}, giving rise to a single Goldstone field $\pi$. Of course, the Lorentz-violating sector could include additional degrees of freedom that may or may not mix with $\pi$ (see \emph{e.g.}\ Ref.~\cite{GaugedGhost} where $\pi$ mixes with a $U(1)$ gauge field).  But from the low energy point of view, it is instructive to see why
this single new degree of freedom is {\it required} in a theory
with gravity by the non-linear realization of broken time
diffeomorphism invariance, independent of any assumptions about the ultraviolet origin of Lorentz violation.

To this end, suppose we begin with the standard model (SM) Lagrangian,
but add a rotationally invariant but Lorentz-violating operator to
the theory
\begin{equation}
\Delta {\cal L} = J_0,
\end{equation}
where $J_\mu$ is a four-vector. There is nothing inconsistent with
doing this in a theory without gravity. But in a gravitational
theory, this operator breaks the time diffeomorphism invariance of
the theory. Now, the diffeomorphism invariance of general
relativity, like all local symmetries, is not a true symmetry but a convenient 
redundancy of description. It serves only to allow us to describe
the two polarizations of a massless graviton in a manifestly
Lorentz invariant way. Starting with the ten degrees of freedom in
the metric fluctuation about flat space $h_{\mu \nu} = g_{\mu \nu}
- \eta_{\mu \nu}$, the diffeomorphism symmetries and their associated constraints eliminate eight
of the degrees of freedom.   The absence of diffeomorphism
invariance is not a conceptual problem, rather it simply means that the
theory is propagating extra degrees of freedom. As familiar from
the physics of gauge symmetry breaking, these new degrees of
freedom are most easily identified by performing a broken symmetry
transformation and elevating the transformation parameters to
fields (see \emph{e.g.} Ref.~\cite{mgrav} for a review).

If we have only broken time diffeomorphisms but leave the spatial
diffeomorphisms intact, we are introducing a single new scalar
degree of freedom $\pi$ into the theory, which can be thought of
as the Goldstone boson for broken time diffeomorphisms (and
Lorentz invariance). An interaction involving this degree of
freedom accompanies all time diffeomorphism- and Lorentz-violating
operators in the theory, with interesting physical consequences.
The low-energy dynamics of $\pi$ is largely determined by
symmetries. In flat space, the dispersion relation for $\pi$ is
forced to be of the form $\omega \propto k^2$. Therefore, sources
moving with any velocity relative to the ``ether" move more
quickly than the velocity of $\pi$ at {\it some}
sufficiently large wavelength, and can therefore emit $\pi$
\v{C}erenkov radiation. The exchange of $\pi$ can give rise to novel
long-range forces. For instance, the leading Lorentz-violating
operator in the SM is the time component of the fermion axial
vector current. This operator is accompanied by a gradient coupling to
$\pi$ which, in the non-relativistic limit, becomes the
spin-gradient coupling $\vec S \cdot \nabla \pi$ familiar from
Goldstone physics. The unusual dispersion relation for $\pi$
implies that the exchange of $\pi$ gives rise to an {\it inverse-square law} spin-dependent force. We will study these phenomena in
detail in the rest of this paper.

\section{Spontaneous Time Diffeomorphism Breaking}

We begin by showing that the presence of Lorentz-violating
terms in the theory necessitates the introduction of a new scalar
degree of freedom with specific couplings largely dictated by
symmetries.  As an example, suppose that in the flat space limit,
our Lagrangian contains a Lorentz-violating term of the form
$J^0$. It is possible to covariantize this term via interactions
with the metric in a way that is invariant under spatial
diffeomorphisms $\xi^i$. If we expand the metric about flat space
$g_{\mu \nu} = \eta_{\mu \nu} + h_{\mu \nu}$, then to leading
order the space-time diffeomorphism generated by $x^\mu
\rightarrow x^\mu + \xi^\mu(x)$ is \be h_{\mu \nu} \rightarrow
h_{\mu \nu} -
\partial_\mu \xi_\nu - \partial_\nu \xi_\mu + \ldots, \ee where
there are additional terms at $\mathcal{O}(\xi^2)$ and
$\mathcal{O}(h \xi)$. To this order, a vector field $J_\mu$
transforms under diffeomorphisms as \be J_\mu \rightarrow J_\mu
- (\partial_\mu \xi^\nu) J_\nu - \xi^\nu
\partial_\nu J_\mu. \ee If we now consider the leading order
couplings between $J_\mu$ and $h_{\mu \nu}$ that break $\xi^0$ but
preserve $\xi^i$ and $SO(3)$ rotations, we find \be
\label{couplingwithhfield} \mathcal{L}_{\mathrm{int}} = \alpha
\left(J_0 + h_{0i} J_i - \frac{1}{2}h_{ii} J_0  \right) + \alpha'
h_{00} J_0 , \ee where $\alpha$ and $\alpha'$ are arbitrary
constants and a sum over $i$ is implied.  Note that the terms in
parenthesis are \emph{forced} to have a shared coupling constant
because of the $\xi^i$ diffeomorphisms.

While the $\xi^i$ diffeomorphisms are exact, the time $\xi^0$
diffeomorphisms are broken, and the theory has one new scalar
degree of freedom. We can isolate this scalar by performing a $\xi^0$
diffeomorphism and elevating $\xi^0$ to a field $\pi$ in a way
that makes the interactions trivially diffeomorphism invariant by
construction. For instance, at linear order in $\xi^0$
\begin{equation}
\left(J_0 + h_{0i} J_i - \frac{1}{2}h_{ii} J_0  \right) \to \left(J_0 + h_{0i} J_i - \frac{1}{2}h_{ii} J_0  \right)
- \vec{J} \cdot \vec{\nabla} \pi,
\end{equation}
which is invariant under $\xi^0$ diffeomorphisms provided that
\begin{equation}
\pi \to \pi - \xi^0.
\end{equation}
The field $\pi$ shifts under time diffeomorphisms and is a new dynamical field
in the theory. Note that the symmetries have forced an interaction
with $\pi$ of the form $\vec{J} \cdot \vec{\nabla} \pi$ to accompany the
Lorentz-violating term $J_0$ in the theory.

So far we haven't described the effective theory for $\pi$ itself.  By the general rules of effective field theory, we should write down all kinetic terms for $\pi$ that are consistent with the unbroken symmetries.  Alternatively, matter loops involving the interaction in equation (\ref{couplingwithhfield}) will generate $\pi$ kinetic terms.  This is in exact analogy with the case of anomalous gauge theories where loops involving the Goldstone coupling $\pi F \tilde{F}$ generate kinetic terms for $\pi$ (and hence a mass term for the anomalous gauge field) \cite{Preskill:1990fr}.

The low-energy Lagrangian for $\pi$ is in fact highly constrained
by the unbroken spatial diffeomorphisms $\xi^i$ and
by the non-linear realization of the $\xi^0$ symmetry, $\pi \to
\pi - \xi^0$. The kinetic term for $\pi$ must be covariantized
by combining it with the metric tensor. We see that the time kinetic term of $\pi$ can easily be
covariantized by combining it with $h_{00}$,
\begin{equation}
\label{eq:time-kinetic} 
\frac12 (\partial_0 \pi)^2 \rightarrow
\frac12 \left(\partial_0 \pi - \frac12 h_{00} \right)^2.
\end{equation}
In unitary gauge where $\pi$ is set to zero by the
diffeomorphism $\xi^0 = \pi$, the time kinetic term for $\pi$
becomes a mass term for $h_{00}$,
\begin{equation}
\frac12 \left(\partial_0 \pi - \frac12 h_{00}\right)^2 \rightarrow \frac18 h_{00}^2.
\end{equation}
On the other hand, the ordinary spatial kinetic term $(\nabla
\pi)^2$ cannot be covariantized with a {\em quadratic} Lagrangian,
because the $h_{0i}$ term which is needed to cancel the time diffeomorphism
of the spatial derivative of $\pi$ also transforms under spatial
diffeomorphisms. The leading spatial kinetic term is $(\nabla^2
\pi)^2$, which comes from the terms
\begin{equation}
\hat{K}_{ij} \hat{K}_{ij}, \qquad \hat{K}_{ii}^2,
\end{equation}
where
\begin{equation}
\label{eq:lin-ext-curv} \hat{K}_{ij}= \frac12(\partial_0 h_{ij}
-\partial_j h_{0i} -\partial_i h_{0j} +2\partial_i \partial_j \pi)
\end{equation}
is the invariant combination under diffeomorphisms.  Therefore, to leading order the kinetic Lagrangian for $\pi$ is
\be
\label{pikinetic}
\mathcal{L}_\mathrm{kinetic} = \frac{1}{2} \dot{\pi}^2 - \frac{1}{2M^2}(\nabla^2 \pi)^2,
\ee
where $M$ is the scale of spontaneous time diffeomorphism breaking and $\pi$ has been canonically normalized.  We readily see the Lorentz-violating dispersion relation for $\pi$, $\omega^2 = k^4/M^2$.  To emphasize, a normal $k^2$ kinetic piece for $\pi$ is \emph{forbidden} by spatial diffeomorphisms. 

To describe the $\pi$ field at the full non-linear level, it is easiest to first go to
unitary gauge by using the time diffeomorphism to set $\pi=0$,
and then to write down terms that are invariant under (time-dependent) spatial diffeomorphism. The basic invariant building
blocks are $g^{00}-1$ and the extrinsic curvature $K_{ij}$ on a
constant-time surface. The three-dimensional curvature $R^{(3)}$
is also invariant but can be expressed in terms of $K_{ij}$ and
the four-dimensional curvature $R^{(4)}$, so it is not
independent. The leading terms at the quadratic order are
\begin{equation}
\label{eq:unitary-gauge} (g^{00}-1)^2,\qquad K_{ij}K_{ij}, \qquad
K_{ii}^2,\qquad (g^{00}-1)K_{ii}.
\end{equation}
The Lagrangian for $\pi$ can then easily be recovered by
performing a time diffeomorphism with $\xi^0=\pi$. This gives rise
the kinetic terms discussed in
equations (\ref{eq:time-kinetic})--(\ref{eq:lin-ext-curv}) at linear
order. Note that an ordinary spatial kinetic term $(\nabla \pi)^2$
is contained in a linear $g^{00}-1$ term at higher order,
\begin{equation}
g^{00}-1 = -h^{00} + h^{0\mu} h_\mu^0 + \cdots \rightarrow (\nabla
\pi)^2,
\end{equation}
where the indices of $h_{\mu\nu}$ are raised and lowered by
$\eta_{\mu\nu}$. However, the $(\nabla \pi)^2$ term is always accompanied by a tadpole
term for the metric, which implies that the space is not in a
ground state. As discussed in Ref.~\cite{GhostOriginal}, the
expansion of the universe will drive the tadpole to zero, and the
coefficient of this term redshifts as $a^{-3}$, where $a$ is the
scale factor. So as long as the redshift starts early enough in
the history of the universe, this term will be so small as to be
totally irrelevant for our discussion.

It is clear that we can equivalently introduce an object $\phi$ as
\begin{equation}
\phi \equiv M^2 t + \pi,
\end{equation}
so that $\phi$ transforms as a usual scalar under diffeomorphisms and also has a separate shift symmetry $\phi \rightarrow \phi + c$.  This was the starting approach in
Ref.~\cite{GhostOriginal}.
It is then convenient to write down Lagrangians involving $\phi$
by following the usual rules of general covariance and then expanding $\phi$ about its time-like vev.  For
instance we can add terms of the form $J_\mu \partial^\mu \phi$ in
the Lagrangian, which become the interaction $M^2 J_0 + J_\mu \partial^\mu \pi$. 
The language of spontaneous time diffeomorphism breaking and the language of $\phi$ getting a time-like vev are completely equivalent as long
as we are only interested in small perturbations around the
vacuum, irrespective of whether $\phi$ is a fundamental scalar
field or a good description far away from the vacuum.

In the presence of the terms from equation (\ref{eq:unitary-gauge}), gravity is
modified: we have a Higgs phase for gravity \cite{GhostOriginal}. 
Among other things, this allows for a new way of having de Sitter phases in the universe, both in the present epoch as 
well as in an earlier inflationary epoch \cite{GhostInflation}. 
A healthy time kinetic term for $\pi$ implies
that the sign of the $h_{00}$ mass term is such that the gravitational
potential is modified in an oscillatory way in the linear regime.
However, non-linear effects are important even for the modest gravitational
sources in the universe such as the Earth. A more detailed analysis of the
non-linear dynamics is discussed in Ref.~\cite{Ghost2}.
The evolution of $\phi$
in a gravitational potential is determined by the local non-linear dynamics
and does not necessarily coincide with that in the cosmological rest frame.
Nevertheless, as long as the $\phi$ background is smooth over some macroscopic
length scale, such as the size of a laboratory or even possibly the size
of the solar system, the spatial dependence of the $\phi$ can be removed
by a local Galilean coordinate transformation, and a local ether rest
frame can be defined. Effectively, this is just redefining what is meant
by the direction and magnitude of the ether ``wind''. This local ether rest
frame is what we refer to in the rest of the paper
when we talk about the velocity with respect to the ether rest frame.

\section{Goldstone Couplings to the Standard Model}

Because we are interested in the $\pi$ dynamics associated with
Lorentz-violation in ordinary matter, we need to know how $\pi$, or
equivalently $\phi$, couples to the standard model fields.
Some couplings are inevitably induced through graviton loops.
In particular, if $\mathcal{O}^{\mu\nu}$ is a symmetric dimension four
standard model operator, then there is no symmetry forbidding the coupling
\begin{equation}
\label{eq:sym-tensor}
\mathcal{L}_{\mathrm{int}} \sim \frac{c}{M^4_{\mathrm{Pl}}}
\mathcal{O}^{\mu \nu} \partial_\mu \phi \partial_\nu \phi \rightarrow c\,
\frac{M^4}{M^4_{\mathrm{Pl}}} \mathcal{O}_{00} - 2c\,
\frac{M^2}{M^4_{\mathrm{Pl}}} \mathcal{O}_{0 i}  \partial_i \pi + \cdots,
\end{equation}
where again $M$ is the mass scale for the Goldstone boson $\pi$,
and the coefficient $c$ is expected to be of order one if this coupling is induced
through gravity.
One such operator is the stress-energy tensor $T^{\mu\nu}$.
The first term in equation (\ref{eq:sym-tensor}) modifies the dispersion
relations for standard model particles. For example, the symmetric
stress-energy tensor for a Dirac fermion is
\begin{equation}
T^{\mu \nu} = \frac{i}{2} \bar{\Psi} (\gamma^\mu \partial^\nu +
\gamma^\nu \partial^\mu) \Psi - \eta^{\mu \nu} \bar{\Psi}
(i \gamma^\rho \partial_\rho - m) \Psi .
\end{equation}
The contribution $(M/M_{\rm Pl})^4 T_{00}$ modifies the dispersion
relation for the fermion to
\begin{equation}
\omega^2  = \left( 1 - c \frac{M^4}{M^4_{\mathrm{Pl}}} \right) (k^2 + m^2).
\end{equation}
This has the effect of changing the maximum attainable velocity for the
fermion. The differences of the maximum attainable velocities for various
particles are constrained to be $\lsim 10^{-21}-10^{-23}$~\cite{Coleman:1998ti}.
This only gives a bound $M \lsim 10^{13}$ GeV if $c$ is $\mathcal{O}(1)$ and is different for different particle species, and this bound is much weaker than
the bound from the nonlinear modification of gravity~\cite{Ghost2}.
Accompanying the modification of the maximum attainable velocity,
the second term in equation (\ref{eq:sym-tensor}) is a coupling of $\pi$
to momentum density. This induces a new velocity-dependent
modification to the Newton's force law. However, due to the $M_{\rm Pl}^4$
suppression, any of these gravity-induced effects are likely be too small to
be observed.

We can also consider direct couplings between $\phi$
and standard model fields which can give larger and potentially observable effects.
To preserve its shift symmetry, $\phi$ must couple derivatively.
The leading coupling to the standard model comes from any dimension
three vector operator $J^\mu$:
\be
\label{piinteractionwithj}
\mathcal{L}_{\mathrm{int}} = \frac{1}{F} J^\mu \partial_\mu \phi
\rightarrow \frac{1}{F}(M^2 J_0 + J^\mu \partial_\mu \pi),
\ee
where $F$ is some unknown mass scale.  Note that this coupling could be forbidden by a $\phi \rightarrow -\phi$
symmetry.  In the Goldstone boson language, this is equivalent to imposing
time reversal invariance in addition to $SO(3)$ invariance.
Nevertheless, this is the leading coupling that could mediate
Lorentz- and CPT-violations to the standard model,
and there is no \emph{a priori} reason to exclude it as long as it satisfies
the experimental constraints.\footnote{
If we do break time reversal invariance, the term $(g^{00}-1)K_{ii} \rightarrow
\dot{\pi} \nabla^2 \pi$ in equation (\ref{eq:unitary-gauge}) is allowed.
The kinetic terms for $\pi$ from equation (\ref{pikinetic}) are slightly
modified and there is an additional $\omega k^2$ term in the $\pi$ dispersion
relation. In this paper, we will ignore this modification because it does
not change the fact that the on-shell condition for $\pi$  still
implies that $\omega \sim k^2$, so at least at a qualitative level,
the Lorentz-violating dynamics will be unchanged.}

The most general vector operator we can create from standard model fermions is
\be
J^\mu = \sum_\psi c_\psi \bar{\psi} \bar{\sigma}^\mu \psi,
\ee
where we have assumed that fermions with the same quantum numbers have been
diagonalized to the $\phi$ interaction basis.
The couplings in equation (\ref{piinteractionwithj}) can actually be
removed via a field redefinition
\be
\psi \rightarrow e^{i c_\psi \phi/ F} \psi,
\ee
but if there are Dirac mass terms or other interactions in the action
that break this $U(1)$ symmetry, then some part of the interaction will
remain.  For concreteness, consider two fermion fields $\psi$ and $\psi^c$
that are joined by a Dirac mass term $m_D \psi \psi^c$.  This mass term
preserves the vector $U(1)$ symmetry but breaks the axial $U(1)$ symmetry,
therefore the coupling to the fermion vector current can be removed but
the coupling to the fermion axial current remains.

In particular, we are left with (in Dirac notation)
\be
\label{piinteractionwithpsi}
\mathcal{L}_{\mathrm{int}} \sim  \frac{1}{F} \bar{\Psi} \gamma^\mu \gamma^5
\Psi \partial_\mu \phi \rightarrow \mu \bar{\Psi} \gamma^0 \gamma^5
\Psi + \frac{1}{F}\bar{\Psi} \gamma^\mu \gamma^5 \Psi \partial_\mu \pi,
\qquad \mu = \frac{M^2}{F}.
\ee
The first term violates Lorentz- and CPT-invariance and gives rise to
different dispersion relations for left- and right-helicity particles
and antiparticles \cite{Kostelecky:1999zh,Andrianov:2001zj},
\be
\omega^2  = (|k| \pm \mu)^2 + m_D^2,
\ee
where the plus sign is for left-helicity particles and antiparticles,
and the minus sign is for right-helicity particles and antiparticles.
Also, if the earth is moving with respect to the ether rest frame, then
after a Lorentz boost, the first term looks like the interaction
\be
\mu \bar{\Psi} \vec{\gamma} \gamma^5 \Psi \cdot \vec{v}_{\mathrm{earth}}.
\ee
In the non-relativistic limit, the current $\bar{\Psi} \vec{\gamma}
\gamma^5 \Psi$ is identified with the spin density $\vec{s}$, giving us
a direct coupling between the velocity of the earth and fermion spin
\be
\mu \vec{s} \cdot \vec{v}_{\mathrm{earth}}.
\ee
Experimental limits on such couplings have placed considerable bounds
on $\mu$.  If we assume that the local ether rest frame is the same
as the rest frame of the CMBR, then
$|\vec{v}_{\mathrm{earth}}| \sim 10^{-3}$.   The bound on couplings to
electrons is $\mu \sim 10^{-25}\GeV$ \cite{Heckel:1999sy} and to nucleons $\mu
\sim  10^{-24}\GeV$ \cite{Phillips:2000dr,Cane:2003wp}. These bounds put limits on the
parameters $M$ and $F$ through the combination  $\mu = M^2/F$.

The second interaction in equation (\ref{piinteractionwithpsi}) leads to
the interesting Lorentz-violating dynamics and will be the focus of the
remainder of the paper.  In the non-relativistic limit
\be
\mathcal{L}_{\mathrm{int}} = \frac{1}{F} \vec{s} \cdot \vec{\nabla} \pi.
\ee
This coupling is familiar from axions, because it is a generic coupling
between fermions and Goldstone bosons.  What makes this different from
the standard story for Goldstone bosons is that $\pi$ has a
Lorentz-violating $\omega \sim k^2/M$ dispersion relation.  The exchange
of a normal Goldstone boson between spin sources leads to a $1/r^3$
spin-dependent potential, but as we will see in
Section~\ref{sectionlongrangespin}, the exchange of $\pi$
leads leads to a $1/r$ potential!  In addition, there is a new
dynamical process that is usually absent in the context of Goldstone
bosons but is familiar from electromagnetism --- ether
\v{C}erenkov radiation. This is due to the fact that given any nonzero
velocity of an object carrying spins, the $\omega \sim k^2$ dispersion
relation for $\pi$ implies that there are always some modes of $\pi$
wave moving slower than the spins. The effects of $\pi$ emission
will be discussed in Section~\ref{sec:emission}.

Before moving on to the discussion of $\pi$ dynamics, we briefly comment
on the Lorentz- and CPT-violating Chern-Simon operator of the electromagnetic
field,
\begin{equation}
\Delta \mathcal{L} = \mu_{\rm CS} \epsilon^{0ijk} A_i F_{jk}.
\end{equation}
This operator is not gauge invariant but gives rise to gauge invariant action
after integration over space-time.  It causes vacuum birefringence and is
strongly constrained by the absence of such effects in radio-astronomy
observations of distant quasars and radio galaxies. The upper bound for
$\mu_{\rm CS}$ is $\mu_{\rm CS} \lsim 10^{-42}$ GeV~\cite{Carroll}, much
stronger than the ones for the axial vector terms of the fermions.
There are some controversies in the literature over whether this term
is generated by the Lorentz- and CPT-violating axial vector term of a
fermion~\cite{Coleman:1998ti, JackiwConcerns, JackiwTwo}. In any case, the absence of divergent contribution
means that there is always a basis where these coefficients are independent
numbers, therefore the strong constraint on $\mu_{\rm CS}$ does not
imply any real constraint on the coefficient $\mu$ of the axial vector
term of the
fermion in equation (\ref{piinteractionwithpsi}). In particular, we can imagine that the reflection symmetry of
$\phi$ is spontaneously broken by the nonzero vacuum expectation value
of another scalar field $S$.  That is, take
\be
\phi \to -\phi, \qquad S\to -S
\ee
 to be a good
symmetry of the theory that is spontaneously broken by $\langle S \rangle$.
Then, the axial vector term for a fermion in fact has to come from
\begin{equation}
\mathcal{L}_\mathrm{int} = S (\partial_\mu \phi) \bar{\Psi} \gamma^\mu \gamma^5 \Psi,
\end{equation}
but the Chern-Simon term cannot be generated because
\begin{equation}
\Delta \mathcal{L} =S (\partial_\mu \phi) \epsilon^{\mu\nu\lambda\kappa} A_\nu F_{\lambda\kappa}
\end{equation}
is not gauge invariant even at the level of the action. Other variations such
as moving the derivative on $\phi$ to another field do not preserve
the shift symmetry on $\phi$. As a result, this vacuum birefringent term
can be forbidden by the gauge invariance and the $\phi$ shift symmetry,
therefore does not give any further constraint on the fermion axial vector
terms.

\section{Ether \v{C}erenkov Radiation}
\label{sec:emission}

In classical electrodynamics, \v{C}erenkov radiation occurs when a charged particle moves through a medium at velocities higher than the speed of light in that medium.  It can be thought of as the optical analog of a sonic boom.  By energy conservation, the charged particle must lose energy in order to generate the photonic shockwave, and once the particle's velocity is less than the medium's light speed, the \v{C}erenkov radiation ceases.

\begin{figure}
\begin{center}
\includegraphics[scale=0.5]{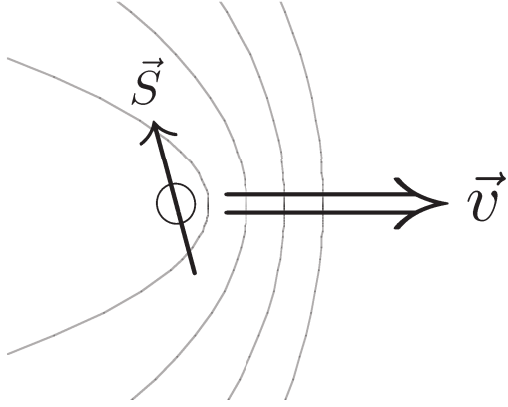}
\end{center}
\caption{A cartoon of ether \v{C}erenkov radiation. The spin $\vec{S}$ is traveling at velocity $\vec{v}$ relative to the ether rest frame. Gray lines are meant to be suggestive of Goldstone shockwaves.}
\label{cherenkovcartoon}
\end{figure}

In the case of our Goldstone boson $\pi$, its dispersion relation is $\omega \sim k^2/M$, so the phase velocity for $\pi$ excitations is
\be
v = \frac{k}{M}.
\ee
For a particle traveling at some fixed velocity, there is always a $k$ such that the speed of the Goldstone is less than the speed of the particle.  As shown in Figure \ref{cherenkovcartoon}, we expect that a particle in motion relative to the ether rest frame --- and which couples to the $\pi$ field --- will radiate away energy until it is at rest with respect to the ether wind.  (For another example of \v{C}erenkov radiation in a different Lorentz-violating context, see Ref. \cite{Lehnert:2004be}.)

With usual \v{C}erenkov radiation, we can use photon detectors to study the photonic shockwave and use that information to understand the motion of the charged particles.  Unfortunately, we do not (yet) have efficient $\pi$ detectors, so the most likely experimental signature of ether \v{C}erenkov radiation would be slight, unexplained kinetic energy loss for particles with spin.  More precisely, depending on the velocity of the observed particle and the velocity of the laboratory frame with respect to the ether rest frame, we would see kinetic energy losses or gains.

We can use a trick to calculate $dE/dt$ for the particle in motion, namely the amount of power needed to maintain the particle's kinetic energy despite the ether drag. The equation of motion for the $\pi$ field in the presence of a spin source is
\be
\label{piequationofmotion}
\ddot{\pi} + \frac{1}{M^2} \nabla^4 \pi = -\frac{1}{F} \vec{\nabla} \cdot \vec{s}.
\ee
Multiplying both sides by $\dot{\pi}$, integrating over all space, and rewriting:
\be
\frac{d}{dt} \left(\int \! d^3 r \; \frac{1}{2}\dot{\pi}^2 + \frac{1}{2 M^2} (\nabla^2 \pi)^2   \right) = \frac{1}{F} \int \! d^3 r \; \vec{s} \cdot \vec{\nabla} \dot{\pi}.
\ee
We recognize the term in parenthesis as the ``particle physics'' energy of the $\pi$ field.  Any energy that goes into the $\pi$ field is energy that we would need to pump into the moving spin to maintain its velocity relative to the ether.  Therefore, the rate of energy loss by the moving spin due to ether \v{C}erenkov radiation is
\be
\label{energydisequation}
\frac{dE_{\mathrm{spin}}}{dt} = - \frac{1}{F} \int \! d^3 r \; \vec{s} \cdot \vec{\nabla} \dot{\pi}.
\ee

We will first calculate this energy dissipation for a spin-density corresponding to a point-like spin moving with velocity $\vec{v}$ relative to the ether:
\be
\vec{s} = \vec{S} \; \delta^{(3)} (\vec{r} - \vec{v}t), \qquad \vec{\tilde{s}} = (2 \pi) \vec{S} \;  \delta (w - \vec{k} \cdot \vec{v}).
\ee
Using Greens functions, the classical solution to equation (\ref{piequationofmotion}) in momentum space is
\be
\tilde{\pi}(\omega, \vec{k}) = \frac{1}{F} \frac{i \vec{k} \cdot \vec{\tilde{s}}}{(\omega + i \epsilon)^2 - k^4/M^2},
\ee
where the $i \epsilon$ ensures that we are using the retarded Greens function.
Plugging this into the energy dissipation formula in equation (\ref{energydisequation}),
\be
\frac{dE_{\mathrm{spin}}}{dt} = \frac{-i M^2}{F^2} \int \frac{k^2dk \, d\Omega_k}{(2\pi)^3} \frac{k^3}{k^2} \frac{(\vec{S} \cdot \hat{k})^2 (\hat{k} \cdot \vec{v})}{(M \hat{k} \cdot \vec{v} + i \epsilon/k)^2 - k^2}.
\ee
At first, it looks like this expression might be zero because it is odd in $k$, but notice there are poles at $k = \pm M \hat{k} \cdot \vec{v} + i \epsilon$.  (We choose our integration ranges such that $\hat{k} \cdot \vec{v}$ is always positive.)  In fact, we see exactly what the poles mean; when the velocity of the source is such that $\pi$ can ``go on-shell'' then there is \v{C}erenkov radiation, and this is the always the case for a non-zero $v$.  Our pole prescription guarantees that the moving particle is radiating $\pi$ energy away to infinity as opposed to receiving $\pi$ radiation from infinity.  We then find
\be
\label{pointspineqn}
\frac{dE_{\mathrm{spin}}}{dt} = - \frac{M^4}{F^2}\frac{|v|}{96 \pi}  \left(|S|^2 |v|^2 + 3 (\vec{S} \cdot \vec{v})^2 \right).
\ee
We see that the rate of energy loss is roughly proportional to $v^3$ and depends on the orientation of the spin with respect to the ether wind.  We can use this result to estimate the expected kinetic energy loss for the most abundant spin point source:  an electron.  Note that we already have a bound on $M^2/F$ of $10^{-25}\GeV$, so if we assume that $|v| \sim 10^{-3}$,
\be
\frac{dE_{\mathrm{electron}}}{dt} \lsim 10^{-37}~\mathrm{GeV\,s}^{-1}.
\ee
This is an incredibly small energy change over a very long period of time, so it is unlikely that we will ever have the experimental precision to track the energy loss of a single electron.

In order to see a measurable effect from ether \v{C}erenkov radiation, we need to have a large value of $S$.  However, it is not enough to simply have a source with a large magnetic moment, as orbital angular momentum generically couples more weakly to $\pi$ than spin.  In particular, the magnetic moment of the earth is not due to spin alignment, so it is not an effective $\pi$ radiator.  Neutron stars are large astrophysical spin sources; a neutron star with the same mass as the sun has a net spin $S \sim 10^{56}$ inside a radius $R \sim 1~\mathrm{km}$.  Closer to home, a 1 ton Alnico magnet has a net spin $S \sim 10^{28}$ inside a radius $R \sim .5~\mathrm{m}$.  Because the spin of these types of objects is spread out over a finite region, we expect the ether drag to be suppressed by some factor of the radius of the source.

\begin{figure}
\begin{center}
\includegraphics[scale=0.8]{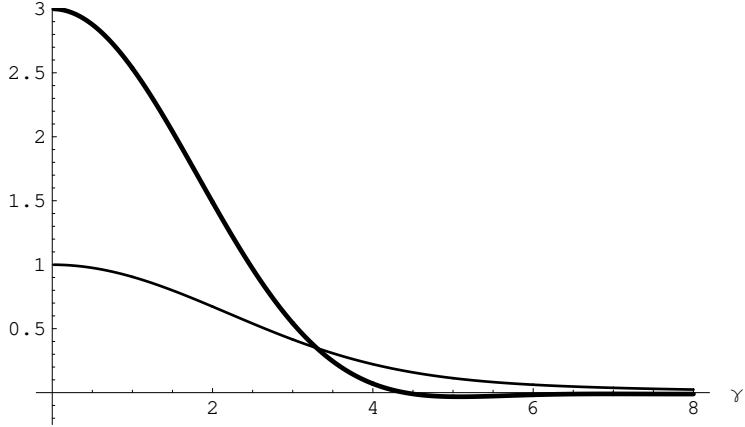}
\end{center}
\caption{The suppression of ether \v{C}erenkov radiation due to finite sources from equation (\ref{rectangleradiation}), $\gamma = MRv$.   The light curve is $Q(\gamma)$ and the bold curve is $R(\gamma)$.  Note that they reproduce the result from equation (\ref{pointspineqn}) when $\gamma \rightarrow 0$.}
\label{cdfunctions}
\end{figure}

For simplicity, consider a rectangle function source:
\be
\vec{s} = \frac{\vec{S}}{\frac{4}{3}\pi R^3} \left\{\begin{array}{ll} 1 & |\vec{r} - \vec{v} t | < R \\ 0 & |\vec{r} - \vec{v} t | > R \end{array}   \right. .
\ee
(We have also considered a Gaussian distribution and the results are nearly identical up to logarithmic factors.)  Following the exact same logic as above, we find
\be
\label{rectangleradiation}
\frac{dE_{\mathrm{rectangle\,spin}}}{dt} = - \frac{M^4}{F^2}\frac{|v|}{96 \pi}  \left( Q(\gamma) |S|^2 |v|^2 + R( \gamma) (\vec{S} \cdot \vec{v})^2\right).
\ee
where $\gamma = M R v$.  Plots of $Q(\gamma)$ and $R(\gamma)$ for small $\gamma$ appear in Figure \ref{cdfunctions}.  For large $\gamma$, the functions behave as
\be
Q(\gamma) \sim \frac{54 \log \gamma}{\gamma^4}, \qquad R(\gamma) \sim -\frac{54 \log \gamma}{\gamma^4}.
\ee
If we assume that $M \sim 10^{-3}\eV$, then the neutron star has $\gamma \sim 5000$ and the 1 ton Alnico magnet has $\gamma \sim 5$.  Na\"{\i}vely applying the bound $M^2/F \sim 10^{-25}\GeV$ for this value of $M$ would give enormous energy losses, but there is an important subtlety in the context of effective field theory.

As we will show in Section \ref{limitssection}, when $\nabla \pi$ becomes large, unknown irrelevant operators dominate the dynamics and our linear analysis breaks down.  Most conservatively, when 
\be
\label{nopredcerenkov}
S \sim F R \gamma^2
\ee
we exit the effective theory in the region of space where \v{C}erenkov radiation dominates.  (We exit the effective theory even sooner in the region where there are no  \v{C}erenkov shockwaves, namely $S \sim F R$.)  For a source with fixed $S$, $R$, and $v$, the maximum value of $dE/dt$ consistent with both the bound $M^2/F \sim 10^{-25}\GeV$ and equation (\ref{nopredcerenkov}) is:
\be
\frac{dE_{\mathrm{max}}}{dt} \sim \frac{S v}{R} 10^{-25} \GeV
\ee
In terms of an anomalous acceleration in the ether rest frame:
\be
a_\mathrm{max} \sim \frac{1}{m v} \frac{dE_{\mathrm{max}}}{dt} \sim \frac{S}{m R} 10^{-25} \GeV.
\ee
For a neutron star and an Alnico magnet, these anomalous accelerations are:
\be
a^{\mathrm{neutron~star}}_\mathrm{max} \sim 10^{-45} \GeV, \qquad a^{\mathrm{magnet}}_\mathrm{max} \sim 
10^{-43} \GeV,
\ee
which, for comparison, is on the order of the Pioneer anomaly ($a_{\rm Pioneer} \sim 10^{-42} \GeV$) \cite{Anderson:1998jd}.  To see this effect for a neutron star seems impossible, unless we were able to map out the matter distribution around the star very precisely.  A real 1 ton Alnico magnet would be fixed to the surface of the earth, so the actual experimental consequence of ether \v{C}erenkov radiation would be a small anomalous force on the earth, but to see it would require measuring a $10^{-9}$ N force on a $1000$ kg object.  At the end of the day, it seems unlikely that we could observe the effects of ether \v{C}erenkov radiation.

We have been focusing on the emission of $\pi$'s from moving spin
sources, but one might ask whether we get strong limits from the cooling of
astrophysical objects, analogous to the supernovae bounds on
axions (see. \emph{e.g.} Ref.~\cite{Raffelt} for a review).  For the supernovae, the relevant temperature is of the
order of $T \sim 30 \MeV$.  For red-giants and the sun the
temperatures are of order $\sim 100 \keV$ to $\sim 1 \keV$. As long
as $T \gg M$, the computation of energy loss to $\pi$'s is outside
the regime of validity of the effective theory and we can't
reliably estimate it. One can imagine that the rate is
extremely small if, for instance, $\pi$ is composite and has a
``size" of order $M^{-1}$, leading to exponentially small
form-factors in its couplings to matter at much smaller distances.
And indeed, in most of the interesting parameter space that we
will consider below, where the long-range forces mediated by $\pi$
exchange are reliably calculable and are of gravitational
strength, we have $M \ll 1 \keV$, so that the astrophysical limits
are inapplicable.

\section{Long-Range Spin-Dependent Potential}
\label{sectionlongrangespin}

The most well known long-range spin-dependent potential is the $1/r^3$ potential transmitted by magnetic fields.  There is also a $1/r^3$ potential coming from pseudoscalar bosons such as axions (see \emph{e.g.} Ref.~\cite{WilczekForces}).   Consider a massless spin-0 field $\varphi$ that has a normal $\omega \sim k$ dispersion relation and a coupling to the fermion axial current, $\vec{s} \cdot \vec{\nabla} \varphi/F$.  In the Born approximation, the potential between two point spins is the Fourier transform of the propagator times the couplings with $\omega \rightarrow 0$.
\be
V_\varphi (r) = \frac{1}{F^2} \int \frac{d^3 k}{(2\pi)^3} \frac{(-i \vec{k} \cdot \vec{S}_1)(-i \vec{k} \cdot \vec{S}_2)}{k^2} e^{i \vec{k}\cdot \vec{r}} = \frac{1}{F^2} (\vec{S}_1 \cdot \vec{\nabla}) (\vec{S}_2 \cdot \vec{\nabla}) \int \frac{d^3 k}{(2\pi)^3} \frac{1}{k^2} e^{i \vec{k}\cdot \vec{r}}
\ee
The Fourier transform of $1/k^2$ is the familiar $1/r$ and we have
\be
\label{axionpotential}
V_\varphi (r) = \frac{1}{F^2} (\vec{S}_1 \cdot \vec{\nabla}) (\vec{S}_2 \cdot \vec{\nabla}) \frac{1}{4 \pi r}  = \frac{-1}{4\pi F^2}\frac{(\vec{S}_1 \cdot \vec{S}_2) - 3 (\vec{S}_1 \cdot \hat{r})  (\vec{S}_2 \cdot \hat{r})}{r^3}
\ee
The form of this potential is identical to the potential between magnetic dipoles in electromagnetism.  Note that the factor of 3 in front of the $(\vec{S}_1 \cdot \hat{r}) (\vec{S}_2 \cdot \hat{r})$ term is directly related to the fact that we have a $1/r^3$ potential.

\begin{figure}
\begin{center}
\includegraphics[scale=0.5]{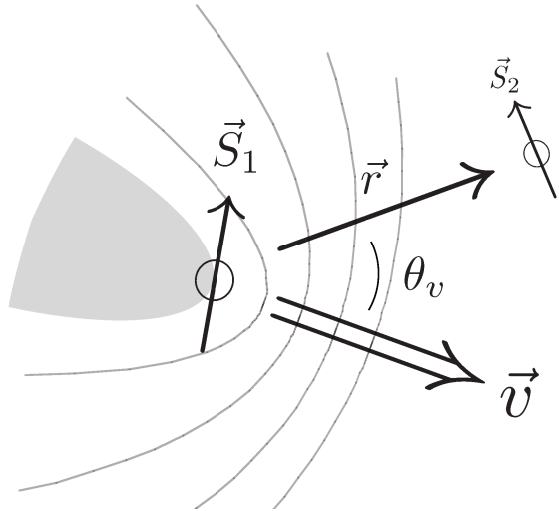}
\end{center}
\caption{A cartoon of the vectors involved in the long-range spin-dependent potential.  The parabolic shadow behind the spin indicates the region where the potential is unsuppressed.  Gray lines are meant to be suggestive of peaks and troughs in the potential.  $\vec{S}_1$ is the source spin moving with velocity $\vec{v}$.  $\vec{S}_2$ is a comoving test spin a distance $r$ away.}
\label{arrangecartoon}
\end{figure}

In the case of $\pi$, we have an $\omega \sim k^2/M$ dispersion relation, so if our sources are in the ether rest frame, the spin-spin potential goes as
\be
\label{notimepotential}
V_\pi (r) = \frac{1}{F^2} (\vec{S}_1 \cdot \vec{\nabla}) (\vec{S}_2 \cdot \vec{\nabla}) \int \frac{d^3 k}{(2\pi)^3} \frac{M^2}{k^4} e^{i \vec{k}\cdot \vec{r}} =  \frac{-M^2}{F^2} (\vec{S}_1 \cdot \vec{\nabla}) (\vec{S}_2 \cdot \vec{\nabla}) \frac{r}{8\pi}.
\ee
Expanding the derivatives:
\be
\label{notimederivativedpotential}
V_\pi (r) =  \frac{-M^2}{8\pi F^2}\frac{(\vec{S}_1 \cdot \vec{S}_2) - (\vec{S}_1 \cdot \hat{r})  (\vec{S}_2 \cdot \hat{r})}{r}.
\ee
The novel dispersion relation for the $\pi$ field has produced a long-range $1/r$ potential between spins!  Assuming that $M/F$ is not too small, we should be able to design experiments to measure this force.  Note that the factor of $1$ in front of the $(\vec{S}_1 \cdot \hat{r}) (\vec{S}_2 \cdot \hat{r})$ term is directly related to the fact that we now have a $1/r$ potential.   Therefore, even if it were impossible to explicitly test the distance dependence of the spin-spin potential, the angular dependence of the potential would still be a smoking gun for a force mediated by spin-0 boson with an $\omega \sim k^2$ dispersion relation.

Searches for anomalous spin-dependent potentials have focused on placing bounds on non-electromagnetic $1/r^3$ potentials (see \emph{e.g.}\ Ref.~\cite{AdelbergerReview} for a review).  Ref.~\cite{Ritter} shows that any new spin-spin $1/r^3$ potential between electrons must be at least a factor of $10^{-13}$ weaker than electromagnetism.  This experiment was carried out at roughly $r \sim 10~\mathrm{cm}$, so na\"{\i}vely comparing the bound to equation (\ref{notimederivativedpotential}),
\be
10^{-13} \frac{\alpha_{EM}}{m_e^2 r^3} \sim \left(\frac{M}{F}\right)^{\!2} \frac{1}{8 \pi r}
\ee 
gives us a limit $M/F \lsim 10^{-19}$.  As we will see in Section $\ref{limitssection}$, this value of $M/F$ is within the interesting range of parameter space and corresponds to forces of gravitational strength.  Of course, the potential mediated by the $\pi$ field has a different angular dependence than the electromagnetic spin-spin potential, so the bound from Ref.~\cite{Ritter} does not apply directly.  However, this does suggest that existing experimental techniques have sufficient precision to begin to probe the Lorentz-violating potential.

In any real experiment we will be dealing with finite sources traveling with some velocity with respect to ether wind, and just like the example of \v{C}erenkov radiation, we expect to see some $\gamma = MRv$ suppression factors in the spin-spin potential.  But even putting that aside, we need to understand what the Born approximation really means in the context of the $\pi$ boson.  By taking $\omega \rightarrow 0$, we are assuming that $\omega \ll k^2/M$.  In position space, this means that our approximation is only valid on time scales
\be
t \gg M r^2.
\ee
For a normal $\omega \sim k$ dispersion relation, we have to only wait a time $t = r$ for our system to behave ``non-relativistically''.  For the $\pi$-mediated forces, however, the time to reach steady state is increased by a factor of $M r$.  For an $M$ of $1\eV$, this factor is $r /(10^{-5}~\mathrm{cm})$ which, given the speed of light, is negligible for any reasonably sized experiment.  However, because we may also want to understand Lorentz-violating dynamics for much larger values of $M$, we want to study how the spin-spin force evolves over time.

Consider a spin source at the origin that turns on at $t=0$:
\be
\vec{s} = \vec{S}_1 \, \delta^{(3)}(\vec{r})\, \theta(t).
\ee
(We have also looked at a spin source that smoothly turns on, and the following results are robust against possible transient effects.) Assuming a test spin $\vec{S}_2$ sitting at $\vec{r}$, the expression for $V_\pi (r, t)$ is
\be
V_\pi (r, t)  =  \frac{1}{F^2} (\vec{S}_1 \cdot \vec{\nabla}) (\vec{S}_2 \cdot \vec{\nabla}) \int \frac{d^3 k}{(2\pi)^3} \frac{M^2}{k^4} \left(1 - \cos (t k^2/M) \right) e^{i \vec{k}\cdot \vec{r}},
\ee
where again we have used a pole prescription corresponding to the retarded potential.  For large $t$, the oscillatory part of the integral vanishes, and we recover the result from equation (\ref{notimepotential}).  When $t = 0$, the potential is zero, as we would expect because information about $\vec{S}_1$ has not yet reached $\vec{S}_2$.  If we had a normal $\omega \sim k$ dispersion relation, then the potential would turn on suddenly when $t = r$.  Here, however, there is no Lorentz invariance, so we have no reason to expect a vanishing potential outside the light-cone.  More precisely, the effective theory breaks down for $k > M$, and because we are not imposing a momentum cutoff, we are inadvertently  propagating modes that travel faster than light.  This effect will disappear for a more realistic turn-on for the source, with time scales much larger than $M^{-1}$.

\begin{figure}
\begin{center}
\includegraphics[scale=0.8]{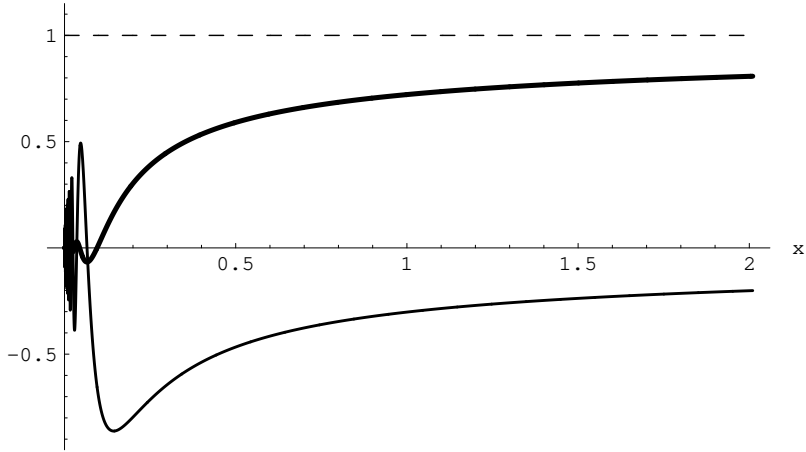}
\end{center}
\caption{The time evolution of the long-range spin-dependent potential from equation (\ref{turnonpotential}), $x = t/Mr^2$.  The bold curve is $K(x)$ and the light curve is $L(x)$.  Note that these functions reproduce the result from equation (\ref{notimederivativedpotential}) in the large $x$ limit.}
\label{klfunctions}
\end{figure}

The time-dependent potential is
\be
\label{turnonpotential}
V_\pi (r, t) = \frac{-M^2}{8 \pi F^2} \left(K(x) \frac{ \vec{S}_1 \cdot \vec{S}_2 - (\vec{S}_1 \cdot \hat{r})  (\vec{S}_2 \cdot \hat{r})}{r} + L(x)\frac{(\vec{S}_1 \cdot \hat{r})  (\vec{S}_2 \cdot \hat{r})}{r} \right),
\ee
where $x = t/ Mr^2$.  Plots of $K(x)$ and $L(x)$ functions appear in Figure \ref{klfunctions}.  We see that the potential does not come to its full value until $t \sim Mr^2$.  For $t$ small compared to $Mr^2$, the potential oscillates between being attractive and repulsive.  The envelope for the oscillations are
\be
|K(x)| \sim 2.3 \sqrt{x}, \qquad |L(x)| \sim 4.5 x^{3/2}.
\ee
Because the potential behaves so erratically for small $x$ values, a realistic experiment will probably have to wait until $x \gg 1$ to see a coherent effect.  In the long time limit
\be
K(x) \sim 1 - \frac{.27}{\sqrt{x}},  \qquad L(x) \sim  - \frac{.27}{\sqrt{x}}.
\ee

Now we consider the effect of finite sources moving with some velocity with respect to the ether rest frame.  In particular, experiments on the earth with magnets fixed to the surface of the earth would be described by sources moving together with a slowly varying velocity $v$.
We might expect that if the source and test spin are traveling fast with respect to the ether, then the $\pi$ waves would not be able to ``keep up'', and the spin-spin potential would be suppressed.  What we will actually find is a far more interesting potential with a striking angular dependence, exhibiting a parabolic shadow behind the spin source, and parabolic shockwaves in front of the spin source.

The finite source is
\be
\label{finitesourcefunction}
\vec{s} = \frac{\vec{S_1}}{\frac{4}{3}\pi R^3} \left\{\begin{array}{ll} 1 & |\vec{r} - \vec{v} t | < R \\ 0 & |\vec{r} - \vec{v} t | > R \end{array}   \right. .
\ee
Following Figure \ref{arrangecartoon}, we want to look at the comoving potential, namely the potential between $\vec{S}_1$ and some test spin $\vec{S}_2$ that is moving at the same velocity as $\vec{S}_1$.
The potential at a comoving distance $r$ is
\be
\label{potentialtocheck}
V(r) = \frac{-M^2}{8\pi F^2} (\vec{S}_1 \cdot \vec{\nabla}) (\vec{S}_2 \cdot \vec{\nabla}) \left( r f(\alpha, \hat{r}\cdot\hat{v},\gamma) \right),
\ee
where $\alpha = M r v$, $\gamma = MRv$, and
\be
f(\alpha, \hat{r}\cdot\hat{v}, \gamma) = \frac{8\pi}{\alpha} \int \frac{d^3 \kappa}{(2\pi)^3} \frac{e^{i \vec{\kappa}\cdot \vec{\alpha}}}{(\vec{\kappa}\cdot \hat{v} + i\epsilon)^2 - \kappa^4} \frac{3( \sin \kappa \gamma - \kappa \gamma \cos \kappa \gamma ) }{(\kappa \gamma)^3}.
\ee
As expected, when $v = 0$ and $R = 0$ then $f(\alpha, \hat{r}\cdot\hat{v}, \gamma) = 1$, and we recover the zero-velocity result from equation (\ref{notimepotential}).  Introducing the notation
\be
\vec{V} \diamond \vec{W} \equiv \vec{V} \cdot \vec{W} - (\vec{V} \cdot \hat{r})(\vec{W} \cdot \hat{r}),
\ee
we can evaluate derivatives on equation (\ref{potentialtocheck}):
\be
\label{velocitysizepotential}
V(r) = \frac{-M^2}{F^2}\frac{1}{8 \pi r} \left(A(\alpha, \theta_v, \gamma) (\vec{S}_1 \diamond \vec{S}_2) + 2 B(\alpha, \theta_v, \gamma) (\vec{S}_1 \cdot \hat{r})(\vec{S}_2 \cdot \hat{r}) \right. \qquad \qquad \qquad \qquad \qquad
\ee
$$
  \qquad \qquad \left. ~+ C(\alpha, \theta_v, \gamma) (\vec{S}_1 \diamond \hat{v})(\vec{S}_2 \diamond \hat{v}) + D(\alpha, \theta_v, \gamma) \left((\vec{S}_1 \diamond \hat{v})(\vec{S}_2 \cdot \hat{r}) + (\vec{S}_1 \cdot \hat{r})(\vec{S}_2 \diamond \hat{v}) \right) \right)
$$
where $\cos \theta_v = \hat{r}\cdot \hat{v}$.  In terms of $f(\alpha, \hat{r}\cdot\hat{v}, \gamma)$,
\be
A(\alpha, \theta_v, \gamma) = F_{00}+F_{10} - F_{01} \cos \theta_v,\qquad B = F_{10}+ F_{20}/2,\qquad 
C = F_{02},\qquad D = F_{11},
\ee
\be
F_{ij} = \alpha^i \left(\frac{\partial}{\partial \alpha} \right)^i \left(\frac{\partial}{\partial ( \hat{r}\cdot \hat{v})} \right)^j f(\alpha, \hat{r}\cdot\hat{v} , \gamma).
\ee
The actual functional forms of $A$, $B$, $C$, and $D$ are not particularly enlightening, so we will evaluate them in certain limits to get an idea of their behavior.  Note that the $A$ component is the only one that is present at zero velocity.

\begin{figure}[t]
\begin{center}
\includegraphics[scale=0.85]{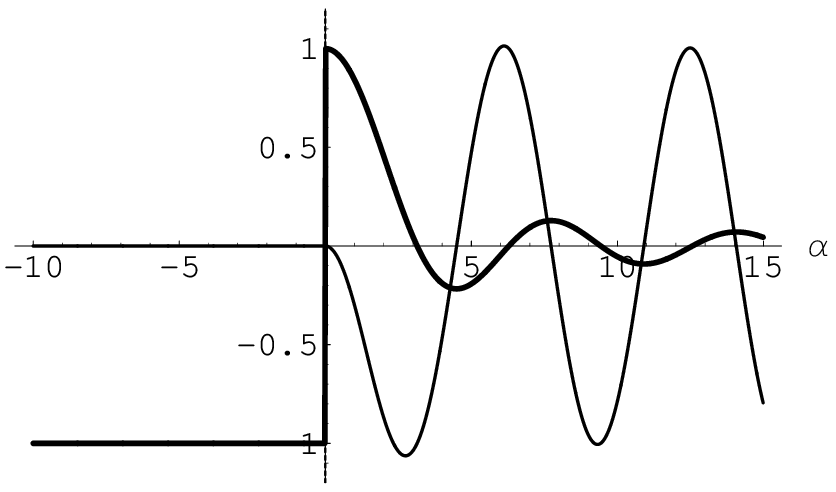} $\qquad$ \includegraphics[scale=0.85]{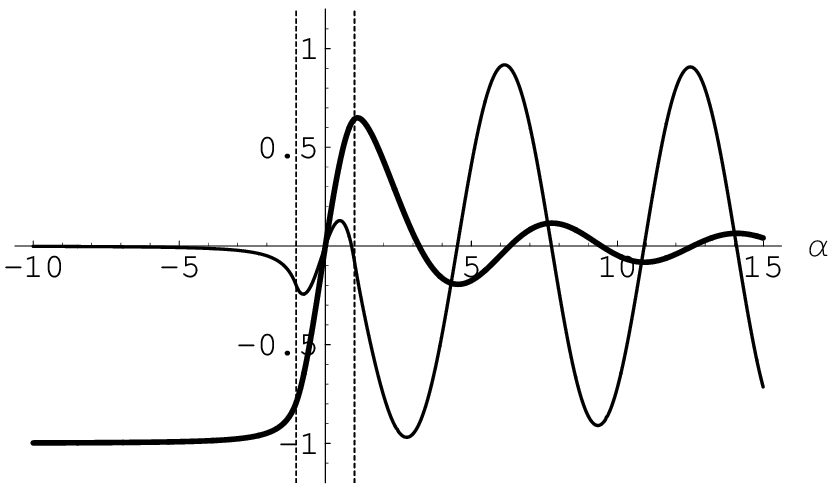}
\end{center}
\begin{center}
\includegraphics[scale=0.85]{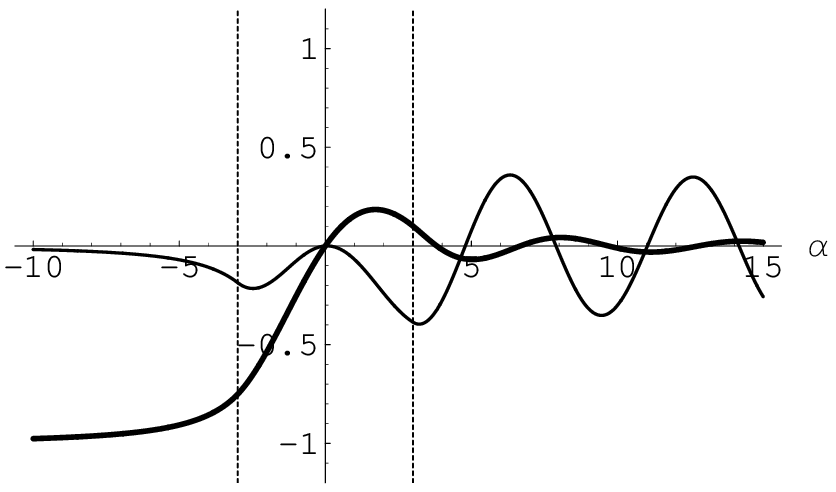} $\qquad$ \includegraphics[scale=0.85]{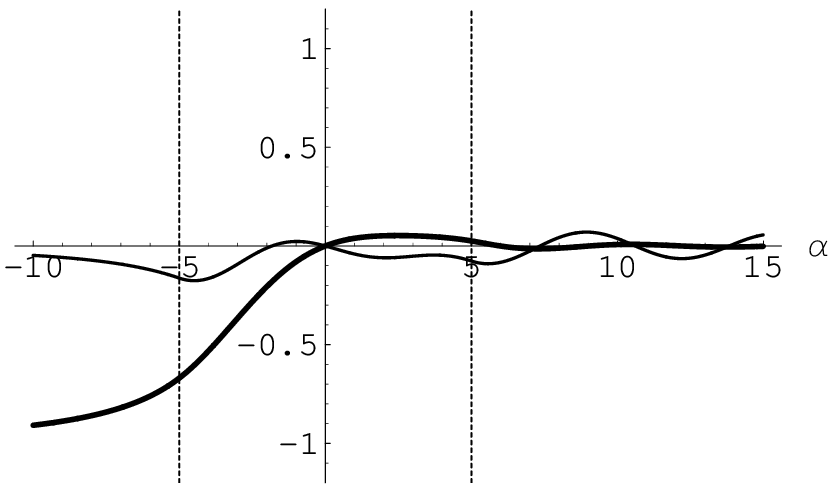}
\end{center}
\caption{The effect of finite sources on the long-range spin-dependent potential from  equation (\ref{velocitysizepotential}), $\alpha = Mrv$ and $\gamma = MRv$.  The bold curves are $A(\alpha, \gamma,\theta_v =0)$ and the light curves are $B(\alpha, \gamma,\theta_v =0)$ for $\gamma =0$, $1$, $2$, and $3$.  Dashed lines indicate the size of the source.}
\label{abfunctions}
\end{figure}

We will start with the case $\theta_v = 0$.  When $\hat{r}$ and $\hat{v}$ are parallel, $\vec{S} \diamond \vec{v} = 0$, so the functions $C$ and $D$ are irrelevant.    When $\gamma = MRv = 0$, there is a nice analytic form for $A$ and $B$:
\be
\label{simpleab}
A(\alpha,0, 0) = \left\{ \begin{array}{ll} -1 & \alpha<0 \\ \sin \alpha / \alpha & \alpha >0 \end{array}   \right., \qquad B(\alpha,0,0) = \left\{ \begin{array}{ll} 0  & \alpha<0 \\ \cos \alpha - \sin \alpha / \alpha & \alpha>0 \end{array}  \right. .
\ee
As $\alpha \rightarrow 0$, we again recover the zero velocity result.  (The minus sign in $A(\alpha <0,0, 0)$ accounts for the fact that we have defined our potential to go as $1/r$, not $1/|r|$.)  At finite velocity, the potential behind the source is the same as the zero velocity potential, whereas in front of the source, the potential oscillates between being attractive and repulsive as $r$ varies!  This confirms our intuition from the existence of ether \v{C}erenkov radiation, because we expect a spin source to leave a potential in its wake.  Behind the source, the $\pi$ field is sufficiently well-established that a test spin does not even know that there is an ether wind present.   In front of the source, the Goldstone shockwaves cannot establish a coherent potential, so we get an oscillating force law.

In Figure \ref{abfunctions} we see the effect of introducing finite sources, namely to ``average'' over the oscillations in equation (\ref{simpleab}).  Far behind the source, we once again recover the zero velocity component of the potential.  For sufficiently small $\gamma = MRv$, the shape of the potential is not significantly modified but the amplitude of the oscillating pieces is somewhat suppressed.  If we consider $M \sim 10^{-3}\eV$, then $1/Mv$ is on the order of tens of centimeters, and we can certainly imagine constructing a spin source such that $\gamma \lsim \mathcal{O}(1)$.  For large $\gamma$, the potential in front of the source is suppressed by $1/\gamma^2$, corresponding to the same $\gamma$ suppression from ether \v{C}erenkov radiation.  Behind the source, the $B$ component vanishes as $1/\gamma^2$, but the $A$ component is unsuppressed, again confirming our intuition that the spin-spin potential should be unmodified from the zero-velocity potential in the ``shadow'' of the spin source.

\begin{figure}
\vspace{-.3in}
\begin{center}
\includegraphics[scale=0.78]{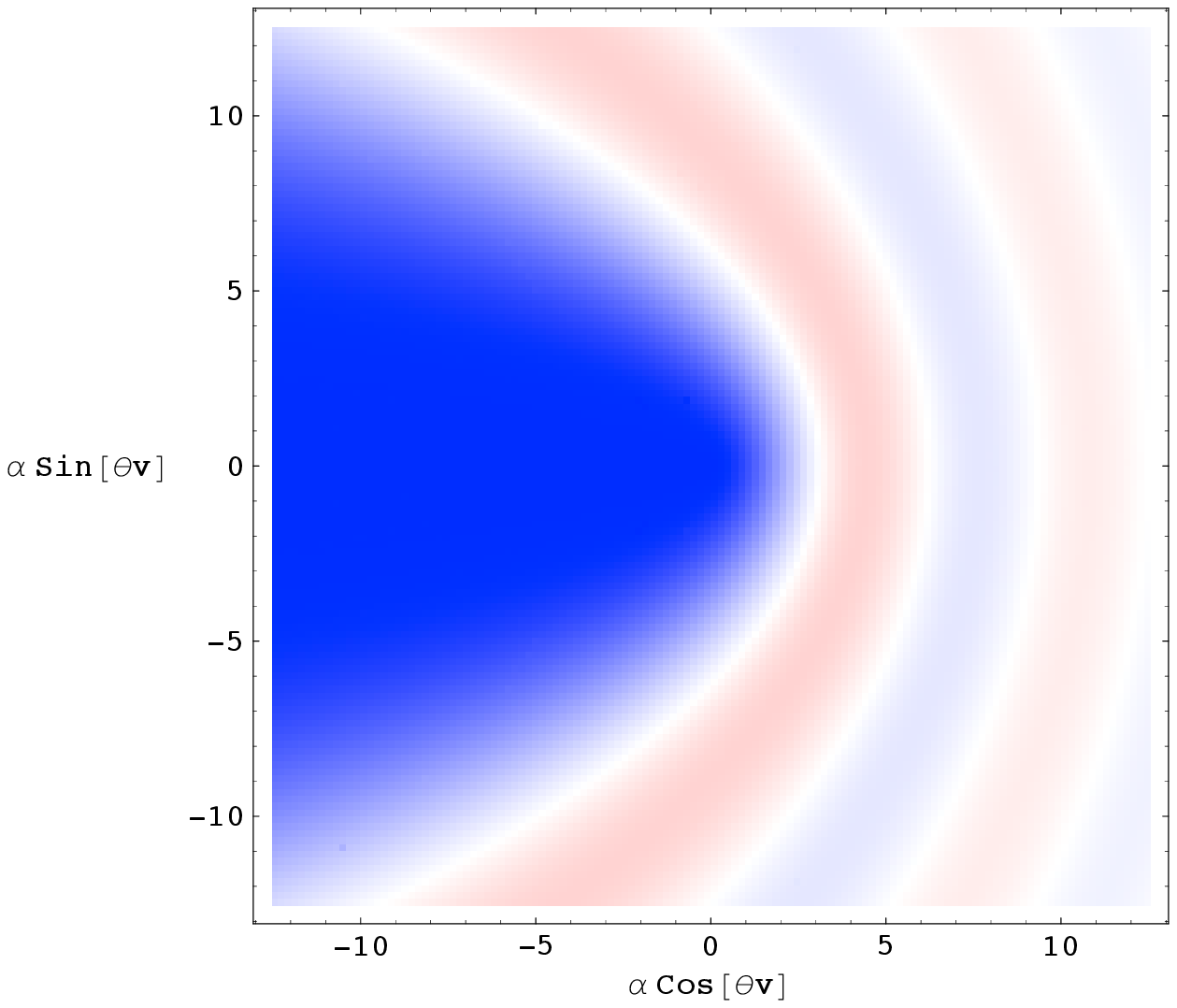}
\end{center}
\vspace{-.6in}
\begin{center}
\includegraphics[scale=0.78]{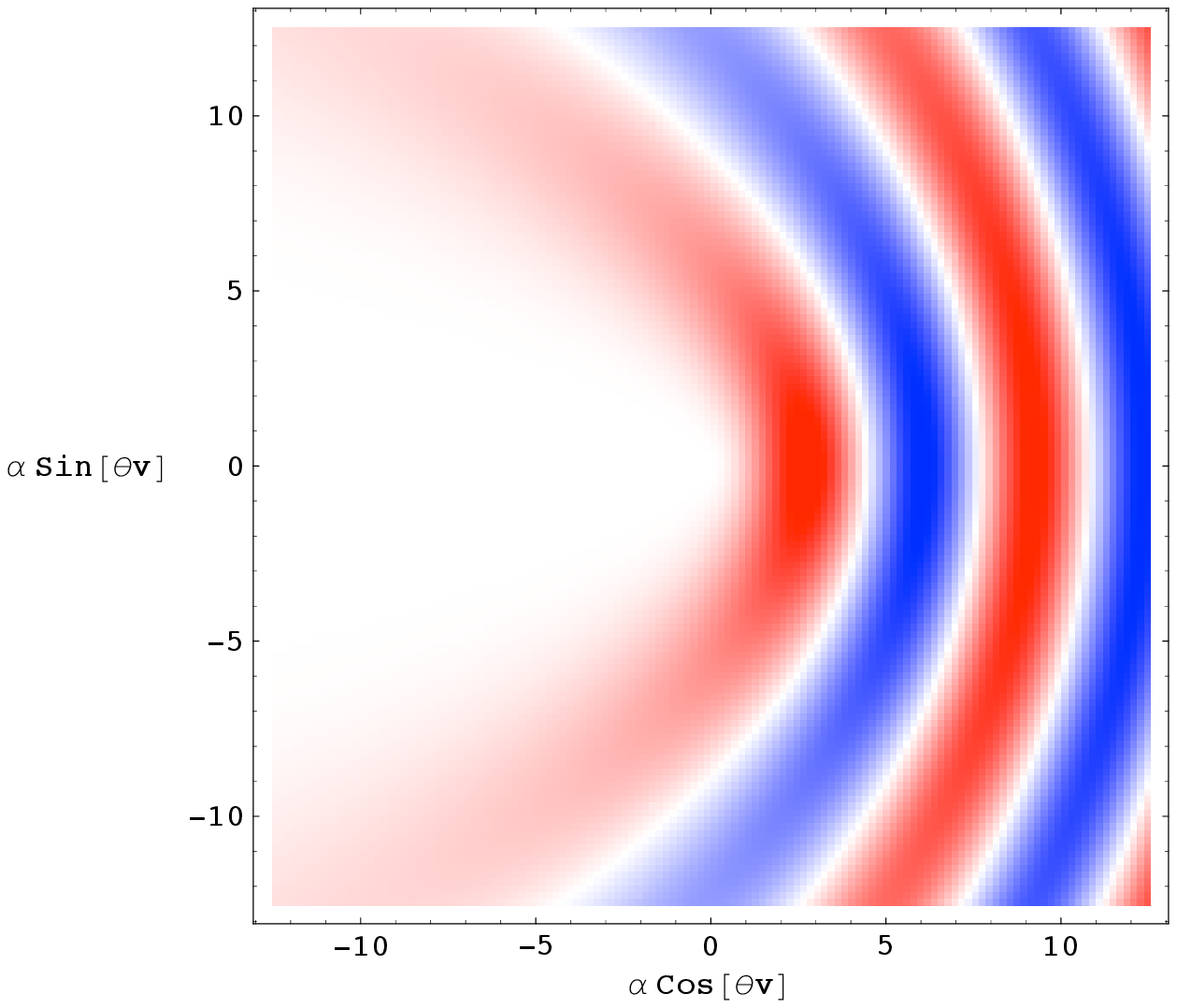}
\end{center}
\vspace{-.3in}
\caption{The angular dependence of the long-range spin dependent potential from equation (\ref{velocitysizepotential}), $\alpha = Mrv$ and $\cos \theta_v = \hat{r} \cdot \hat{v}$.  The top function is $A(\alpha, \gamma = 0, \theta_v)$ and the bottom function is $B(\alpha, \gamma = 0, \theta_v)$.  Blue indicates positive potential, red indicates negative potential, and white indicates zero potential.  The parabolic shape of the potential peaks and troughs are a direct consequence of the $\omega \sim k^2/M$ dispersion relation for the $\pi$ field.  The parabolic shadow in the $A$ component shows that behind the spin source, a test spin feels the zero-velocity potential.  It is also intuitively obvious that the ``ether wind'' is blowing to the left.}
\label{abfunctionangle}
\end{figure}

 If we assume that $R = 0$, then the most spectacular prediction of $\pi$-mediated spin potentials is the angular dependence.   In Figure \ref{abfunctionangle}, we see the value of the $A$ and $B$ components of the potential as a function of $\alpha$ and $\theta_v$.  (The $C$ and $D$ components look roughly similar to the $B$ component.)  The parabolic shape of the potential oscillations is a direct consequence of the $\omega \propto k^2$ dispersion relation for the $\pi$ field and are a sharp prediction of $\pi$-mediated forces.  In the $A$ (zero-velocity) component, we see the parabolic shadow cast by the spin source.  In a realistic situation, both the $A$ and $B$ components would vanish for sufficiently large $\alpha$ because as we saw in Figure \ref{klfunctions}, it takes a time $t \sim Mr^2$ for the potential to reach steady state.  
 
By mapping out the potential for various orientations of the spins and various values of $\theta_v$, it would be possible to see the $\pi$ field and determine the direction of the ether wind and the value of $Mv$.  Even if our spin source and test spin were fixed to the surface of the earth, the value of $\theta_v$ would still vary over the course of a day!  As we will see in Section \ref{limitssection}, the magnitude of the spin-dependent potential is generically weaker than gravity, but because the angular dependence is so different from gravity or even magnetism, it may be possible to extract the $\pi$ component of any $1/r$ potential, because as is evident from Figure \ref{abfunctionangle}, the $\pi$ potential defines a preferred axis in space.

We have seen that the scale $M$ sets the characteristic length scale and the characteristic time scale for the spin-spin potential.  In the next section, we will find that the potential can be of gravitational strength for $M$ between $10^{-3} \eV$ and $10 \eV$.  For these scales, the retardation effects from Figure \ref{klfunctions} are completely negligible.  The scale $1/Mv$ is between fractions of a millimeter to tens of centimeters, so if one wants to capture all of the oscillating pieces of the potential from Figure \ref{abfunctions}, an ideal experimental spin-source would be millimeter- or centimeter-sized in order to avoid the $1/\gamma^2$ size-suppression effects on the oscillations.  Similarly, the separation between spins should be roughly in the $1~\mathrm{cm}$ to $10~\mathrm{cm}$ range.  These scales suggest that the long-range spin-dependent potential should be accessible to precision tabletop experiments.   However, if one is only interested in seeing the parabolic shadow of Figure \ref{abfunctionangle} and not the fully oscillatory structure, then the size $\gamma$ of the spin source and the separation $\alpha$ can be arbitrarily large.    

Note that this energy range for $M$ is also interesting cosmologically.   If $M \sim 10^{-3}~\mathrm{eV}$, then naturalness suggests that the Goldstone sector could explain the observed acceleration of the universe even if the cosmological constant were zero.  Alternatively, $M \sim 1~\mathrm{eV}$ is around the temperature of matter domination in our universe, perhaps indicating that the $\pi$ boson is a dark matter candidate.  Of course, there is no reason why $M$ could not be much larger than these scales, but it is interesting that the range for $M$ that is experimentally accessible is also cosmologically relevant.  In fact, for $M \sim 10^{-3}~\mathrm{eV}$, no cosmological experiment could distinguish between a cosmological constant and a Goldstone sector \cite{GhostOriginal}, and the \emph{only} way we would know about the Goldstone sector would be through direct couplings to the standard model.  

\section{Limits on the Effective Theory}

\label{limitssection}

As a final check that our analysis makes sense, we need to verify that in the presence of large sources, the $\pi$ field is still in the range of validity for its effective theory.  In particular, when
\be
\label{limitnopredict}
\nabla \pi \sim M^2,
\ee
our theory loses predictability because irrelevant operators make large, unknown contributions to the action.    This is not the only concern, however.  Even if the theory has predictability, large sources might push us out of the linear regime. The least irrelevant interaction for the $\pi$ field in the static limit ($\dot{\pi} = 0$) is $(\nabla \pi)^4/M^4$, and if this term is larger than the source term $\vec{s} \cdot \vec{\nabla} \pi / F$, then self-interactions dominate over sources.  In particular, $|\vec{s}| \sim S/R^3$ so non-linear effects become important when
\be
\label{limitnonlinear}
\frac{(\nabla \pi)^3}{M^4} \sim \frac{S}{F R^3}.
\ee
In this case, while the effective theory may still be well-behaved, our linearized analysis is no longer valid because we have ignored non-linear terms in the equation of motion for the $\pi$ field.  Of course, these non-linear effects are in principle tractable, but we will postpone a full non-linear analysis.

One subtlety is which choice of $\nabla \pi$ we should use in equations (\ref{limitnopredict}) and (\ref{limitnonlinear}).  The long range spin potential was dominated by $\nabla \pi$ in the shadow of the spin source, but ether \v{C}erenkov radiation has to do with Goldstone shockwaves appearing in front of the the spin source.  While we could certainly claim that the relevant value of $\nabla \pi$ is just the maximum value over all space, to be conservative, we will cite separate bounds for $\nabla \pi$ in the shadow and from the shockwaves.  This corresponds to the expectation that even if $\pi$ is, say, deep in the nonlinear regime in one region, we can still use a linear analysis if another region dominates the dynamics.  

We can easily calculate the magnitude of $\nabla \pi$ for the source in equation (\ref{finitesourcefunction}).  Note that $\nabla \pi$ in a comoving frame is the same as $\nabla \pi$ is the ether rest frame, because in the non-relativistic limit the two frames are connected by a Galilean transformation which does not affect spatial derivatives.  For large finite sources, the maximum value of $\nabla \pi$ in the shadow occurs directly behind the spin source, and in the shockwave region, the maximum value of $\nabla \pi$ occurs directly in front of the spin source and is suppressed relative to the shadow value:
\be
|\nabla \pi|_{\rm shadow} \sim \frac{S M^2}{FR}, \qquad |\nabla \pi|_{\rm shockwave} \sim \frac{S M^2}{FR} \frac{1}{\gamma^2},
\ee
where $R$ is the radius of the source and $\gamma = MRv$.  The bounds on the size of the source are
$$
S^{\mathrm{no\,predictability}}_{\rm shadow} \sim FR, \qquad S^{\mathrm{non-linear}}_{\rm shadow} \sim \frac{F}{M},
$$
\be
\label{boundsons}
S^{\mathrm{no\,predictability}}_{\rm shockwave} \sim FR \gamma^2, \qquad S^{\mathrm{non-linear}}_{\rm shockwave} \sim \frac{F}{M} \gamma^3.
\ee

Note that just because there may exist a spin source that violates the bounds on the effective theory, it does not mean that  we can use this information to place constraints on $M$ and $F$.  In our entire analysis, we are assuming that full diffeomorphism invariance is restored in the UV, so the irrelevant interactions of the $\pi$ field must somehow encode the fact that Lorentz symmetry is actually a good symmetry of the complete theory.  Therefore, though we cannot trust the theory of $\pi$ bosons around large sources, the fact that large sources exist does not mean that the $\pi$ description is not valid around smaller sources.

\begin{figure}[p]
\begin{center}
\includegraphics[scale=0.56]{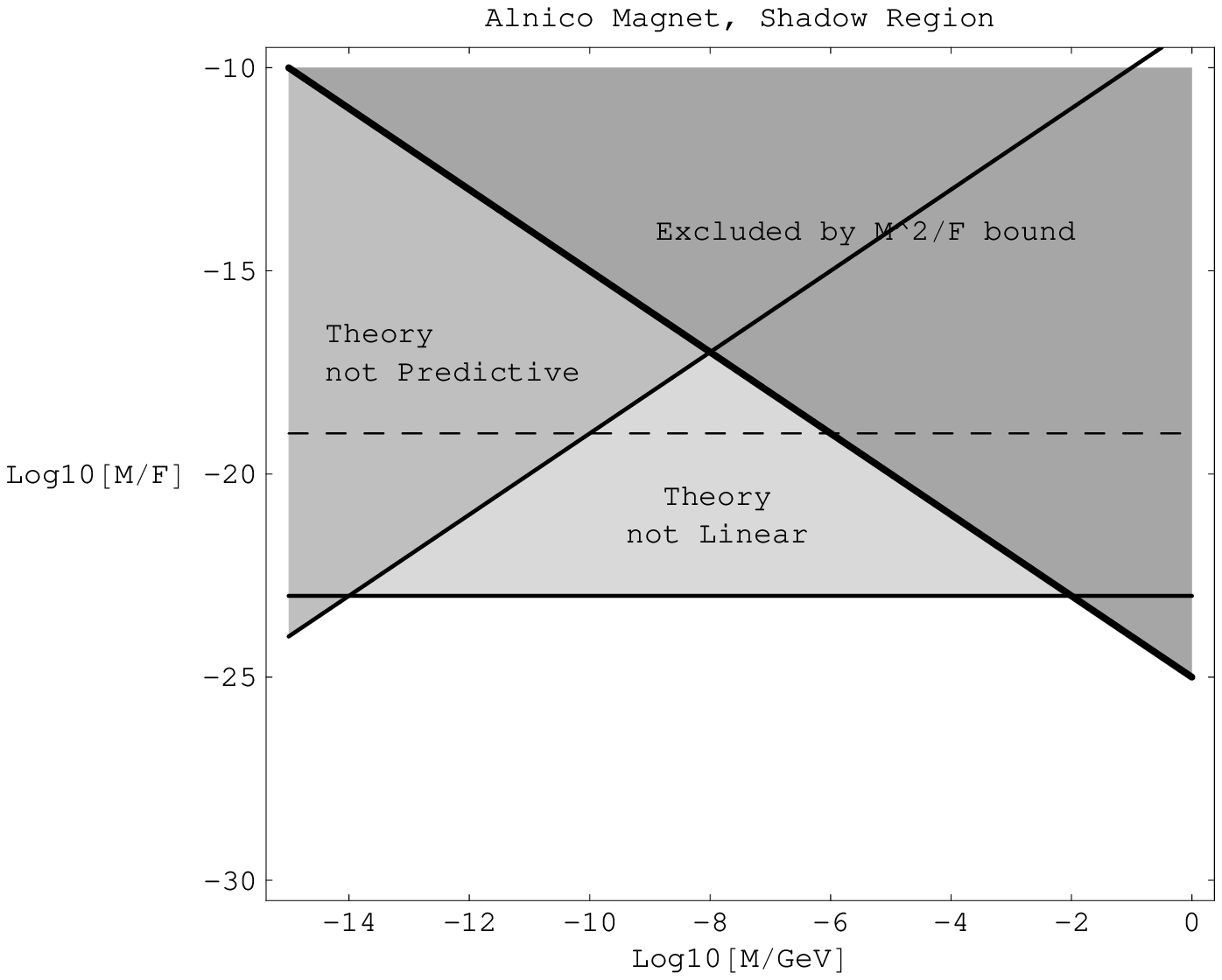} \quad \includegraphics[scale=0.56]{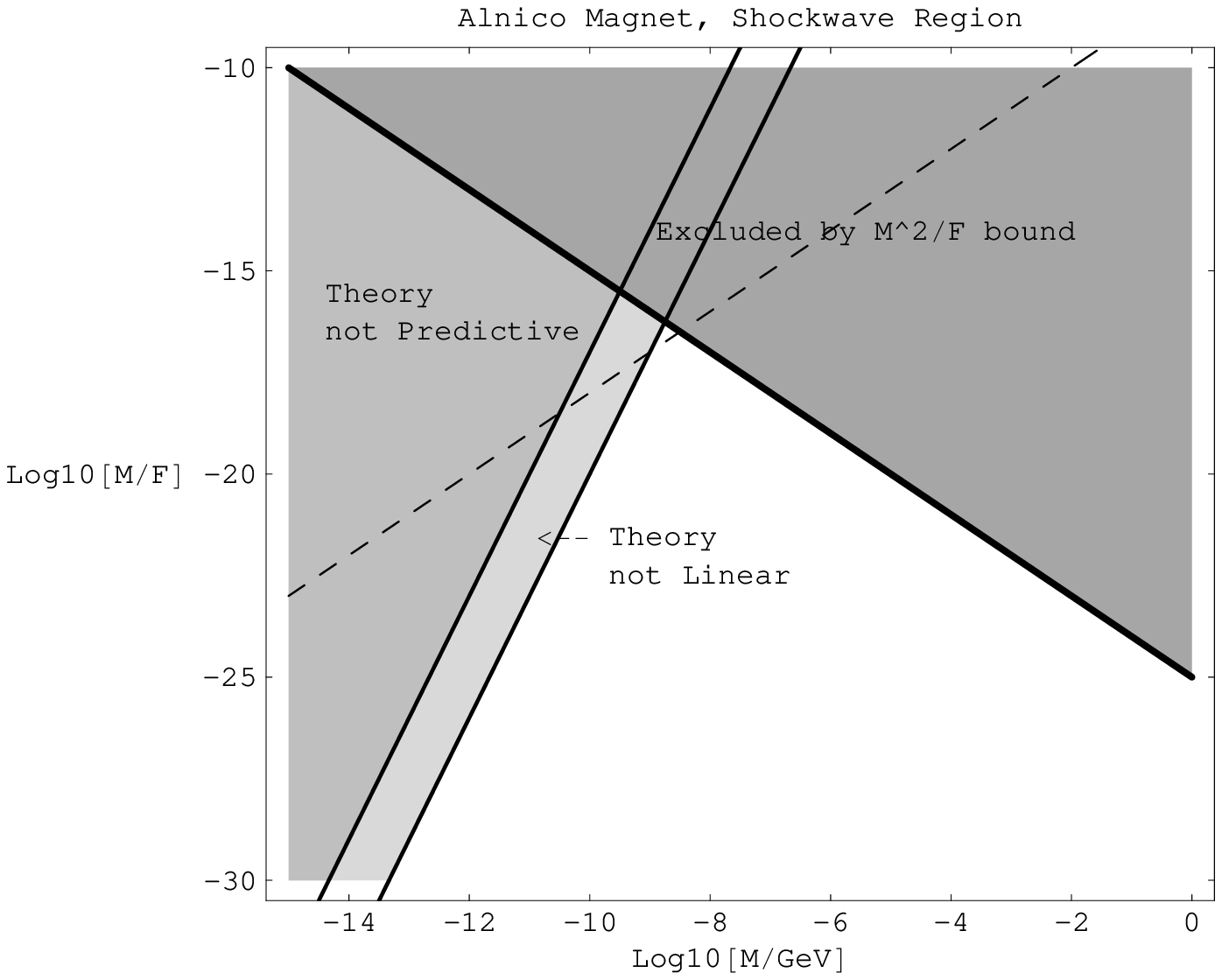}
\end{center}
\caption{Bounds on the effective theory of $\pi$ bosons assuming an Alnico test spin of radius $R \sim 1~{\rm cm}$.  The dashed line represents ``gravitational strength,'' namely when the $\pi$-mediated force is the same strength as gravity assuming one aligned spin per nucleon.  Precision tabletop experiments are sensitive to forces of sub-gravitational strength, so the fact that the dashed line is within the nonlinear regime does not automatically exclude direct detection for smaller values of $M/F$. The experimental bound on $M^2/F$ is $10^{-25}\GeV$.}
\label{boundsalnico}
\end{figure}

\begin{figure}[p]
\begin{center}
\includegraphics[scale=0.56]{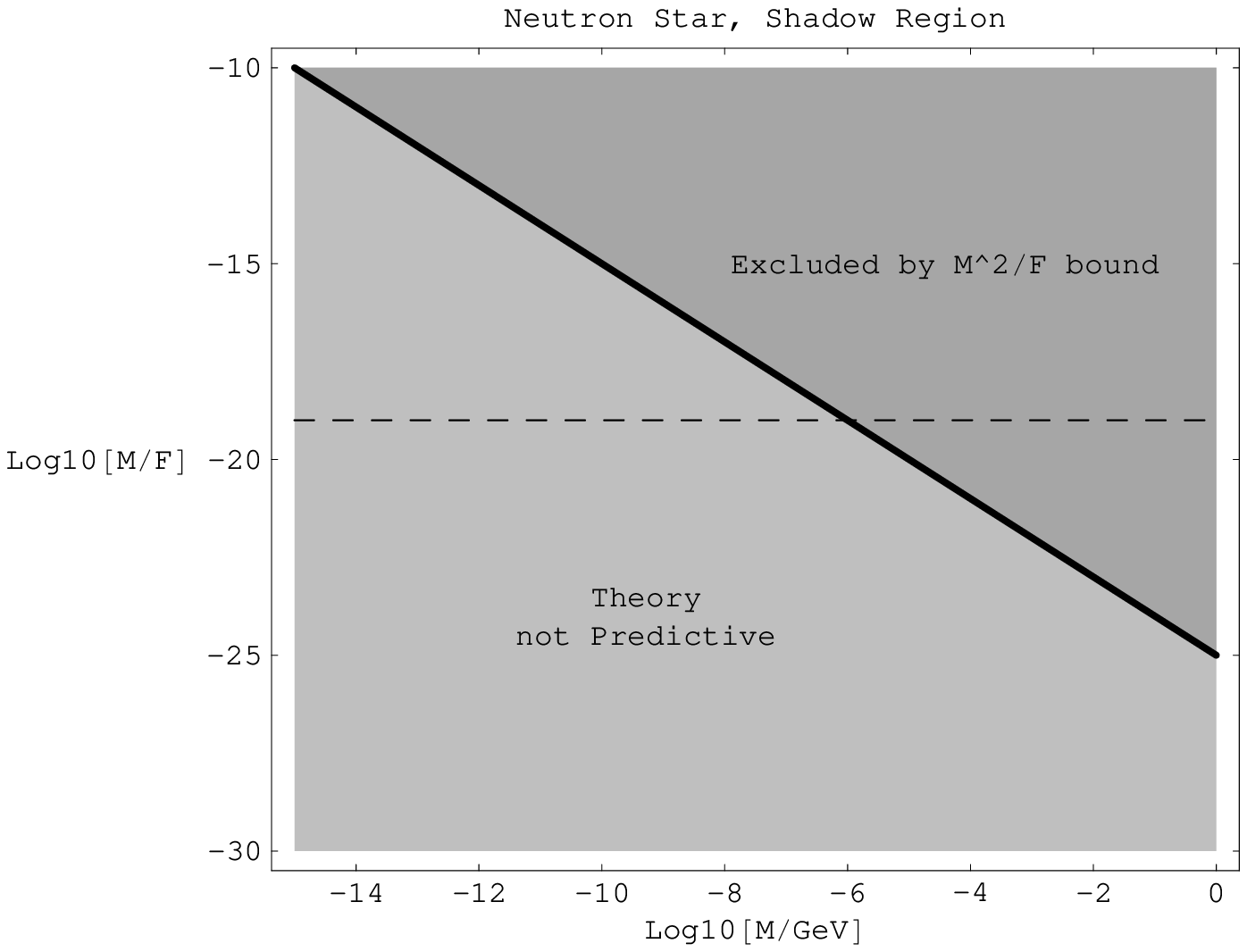} \quad 
\includegraphics[scale=0.56]{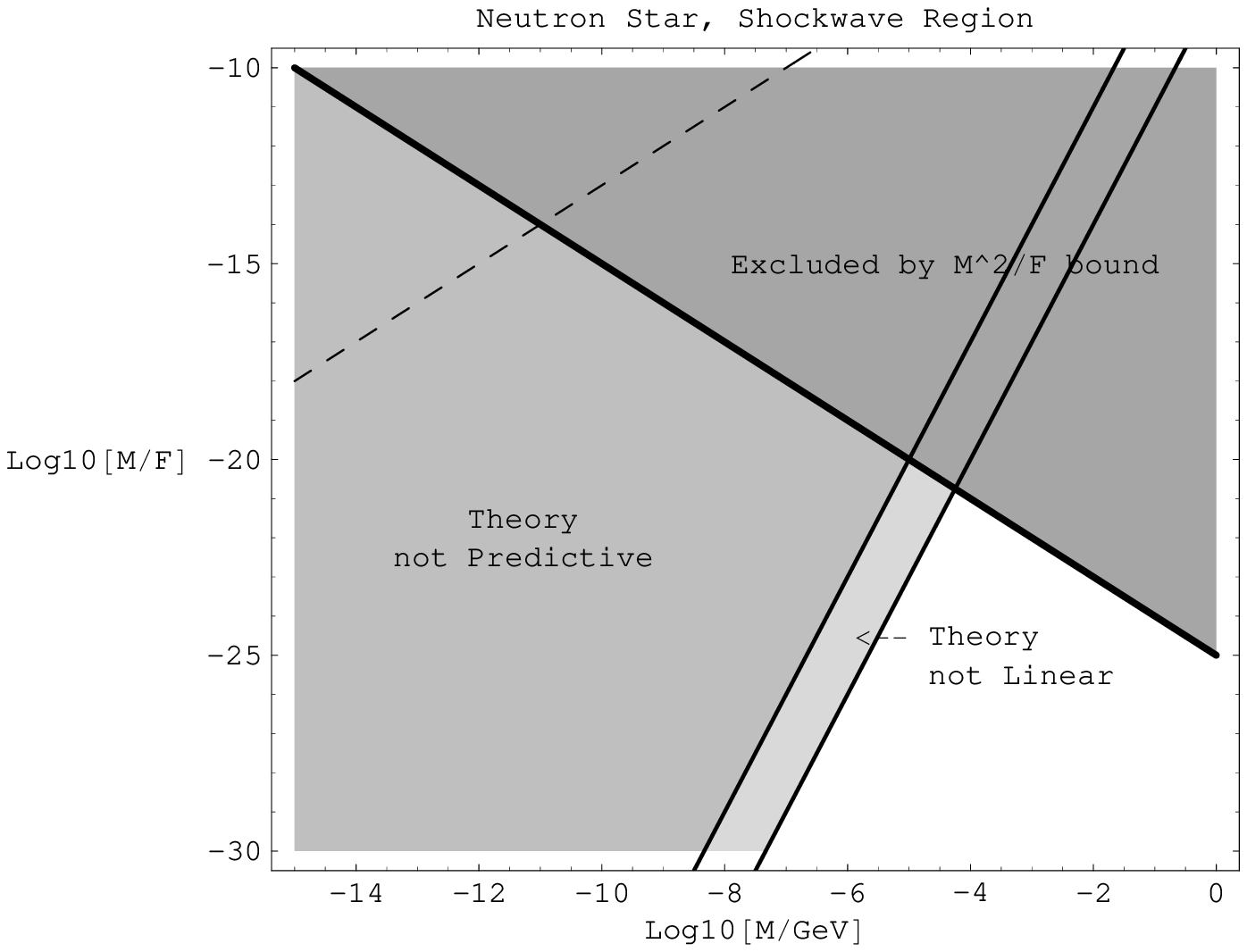}
\end{center}
\caption{Bounds on the effective theory of $\pi$ bosons assuming a neutron star wtih $S \sim 10^{56}$ and $R \sim 1~\mathrm{km}$.  The dashed line represents ``gravitational strength'' as in Figure \ref{boundsalnico}.  Here, we are implicitly assuming that the $\pi$ field couples to both nucleons and electrons, but it may be the case that couplings to nucleons are suppressed or even entirely absent.}
\label{boundsstar}
\end{figure}

We are now ready to see the phenomenologically viable and experimentally testable regions of our parameter space.  The strength of the long-range spin-dependent potential goes as $(M/F)^2$, so it is convenient to place bounds in terms of $M/F$ and $M$.  If we assume a source with one aligned spin per nucleon, then the spin-spin potential is of gravitational strength when
\be
\frac{M}{F} \sim \frac{m_{\rm nucleon}}{M_\mathrm{Pl}} \sim 10^{-19} \quad (\mbox{shadow region}), \qquad \frac{M}{F} \sim 10^{-19} \gamma \quad (\mbox{shockwave region}),
\ee
where the factor of $\gamma$ accounts for the $1/\gamma^2$ suppression of the spin-spin potential in front of the source.  We will use an Alnico magnet as our canonical spin source:
\be
S \sim R^3 \left(\frac{10^{23}~\mathrm{spins}}{\mathrm{cm}^3}  \right).
\ee
If we assume that the magnet is moving relative to the ether with $v \sim 10^{-3}$, then the relevant bounds in terms of $R$ are:
\be
\begin{tabular}{l|ll} & To be Predictive &  To be Linear \\
\hline
\tabwide Shadow & ${ \displaystyle \frac{M}{F} < \left(\frac{\rm cm}{R} \right)^2 \frac{M}{10^9 \GeV}}$& ${\displaystyle \frac{M}{F}< 10^{-23} \left(\frac{\rm cm}{R} \right)^3}$ \\
\tabwide Shockwave & ${\displaystyle \frac{M}{F}< \left(  \frac{M}{10^{-4} \GeV}  \right)^3}$ & ${\displaystyle \frac{M}{F}< \left(  \frac{M}{10^{-3} \GeV}  \right)^3}$
\end{tabular}
\ee
Combining these bounds with the experimental bound  $M^2/F \lsim 10^{-25}\GeV$, Figure \ref{boundsalnico} shows the experimentally testable region for an Alnico spin source with $R \sim 1~{\rm cm}$.
In both the shockwave and shadow regions, non-linear effects are important when the force is of gravitational strength, but in the shadow region, we can reduce the effect of non-linearities simply by decreasing the total spin of our magnet.  Also, precision tabletop experiments are usually sensitive to sub-gravitational forces, so there is a possibility for direct detection with smaller values of $M/F$.  We see that there are indeed regions of parameters space where the force is some reasonable fraction of gravitational strength, where our theory is within the linear regime, and which are not excluded by experimental bounds.   In this region, $M$ is between around $10^{-3} \eV$ and $10 \eV$, though unknown non-linear effects become increasingly important as we increase the size of our spin source.

For comparison, the bounds for a neutron star with the same mass of the sun traveling at $v \sim 10^{-3}$ with respect to the ether rest frame appear in Figure \ref{boundsstar}.  Again, the lower shaded regions in this figure are not regions that are excluded by experiment but simply regions where our linear analysis has no predictive value.  Also, if for some reason $\pi$ coupled to electrons but did not couple to nucleons, then the neutron star bounds are absent.

\section{Conclusions}

Lorentz invariance has been one of the foundations of modern
physics for the past century, so the discovery of spontaneous
Lorentz violation would be very exciting. So far, Lorentz
invariance has survived all of the extremely accurate experimental
and observational tests to date.  It will surely be continuously tested by even
more precise experiments in the future.

If Lorentz invariance is broken, there must be a Goldstone mode
associated with it. Because violations of Lorentz invariance also
imply the breakdown of general covariance, there are not
enough diffeomorphism symmetries to reduce the number of propagating degrees
of freedom in the metric tensor down to the usual two
polarizations of the graviton. In unitary gauge, the extra
degree of freedom is precisely the eaten Goldstone mode. The
Lagrangian of the Goldstone mode can be determined by the
remaining spatial diffeomorphism symmetries and the stability of
the vacuum. In particular, for the case of Lorentz-violation in
the time direction, the kinetic term for the corresponding
Goldstone boson was determined by the diffeomorphism symmetry to be
\be
\mathcal{L}_\mathrm{kinetic} = \frac12 \dot{\pi}^2-\frac{1}{2M^2} (\nabla^2 \pi)^2,
\ee
which gives
rise to the novel Lorentz-violating dispersion relation $\omega^2 \sim k^4/M^2$.
The interactions of the Goldstone boson with ordinary matter
fields can also be derived by performing a broken time
diffeomorphism transformation on the corresponding
Lorentz-violating term in the Lagrangian. These interactions
provide additional tests of Lorentz violation complementary to
traditional tests looking for the presence of explicit Lorentz- and CPT-violating operators.
Also, if the scale $M~\sim~10^{-3} \eV$ sets the cosmological constant, then the only way we could probe the Goldstone sector is through direct couplings to the standard model.

The possible effects of these new interactions include emission
and exchange of the Goldstone boson. The leading interaction is
expected to be the coupling to the axial vector of standard
model fermions, which reduces to spin in the non-relativistic
limit. For emission, a potential bound would come from star
cooling similar to the usual axion constraint, though we have argued that such constraints may not be relevant if $M$ is smaller than the typical temperatures of astrophysical objects. Another interesting
effect is the \v{C}erenkov radiation of the Goldstone boson.
Because of the $\omega^2 = k^4/M^2$ dispersion relation, for a
source moving relative to the ether rest frame, there are always
modes with long wavelengths travelling slower than the velocity of
the source, and therefore ether \v{C}erenkov radiation will be
emitted. Because there is no fixed speed for Goldstone waves, the
\v{C}erenkov radiation is not emitted at a fixed angle. The
wavefronts look like parabolas rather than straight lines.
Unfortunately, if we restrict ourself in the regime of validity of
the effective theory where these effects are under theoretical
control, the energy loss due to the emission of the Goldstone
bosons are too small to give promising tests of Lorentz violation
beyond the current direct constraints.

Exchange of the Goldstone boson, on the other hand, induces new
forces between spin sources. In the static limit, the $1/k^4$
propagator gives rise to $1/r$ potential between spin sources
instead of the usual $1/r^3$ potential from a $1/k^2$ propagator.
This is a distinct signature of such a Lorentz-violating
propagator. In a laboratory, however, we do not expect that the
spin sources are at rest respect to the ether rest frame. The
potential is then modified by the velocity effect, generating a distinct parabolic
potential with a shadow behind the source and shockwaves in front of the source. Such a potential may be observable in regions of
the parameter space that are both allowed by the direct Lorentz-violation
constraints and where the effective theory is still under theoretical
control.  We expect the strongest signal to be of gravitational strength, coming from millimeter- to centimeter-sized spin sources separated by centimeter distances.  Intriguingly, this is right around the sensitivity of current experiments.  If observed in future experiments, this would be a
spectacular signal for spontaneous Lorentz violation.

{\bf Acknowledgments}: We wish to thank Devin Walker and Ralf Lehnert for useful discussions.

 \end{document}